\documentclass[preprint,prd]{revtex4-1}
\usepackage{color}
\usepackage{graphicx}
\usepackage{epsfig}
\usepackage{subfig}

\usepackage{graphicx}
\usepackage{dcolumn}
\usepackage{bm}

\begin{document}

\title{Electromagnetic Duality  and New Solutions
 of  the Non-minimally Coupled $Y(R)$-Maxwell Gravity}

\author{ \"{O}zcan SERT}
\email{osert@pau.edu.tr}
 \affiliation{Department of Mathematics, Faculty of Arts and Sciences,
Pamukkale
University,  20017   Denizli, T\"{u}rkiye}


\date{\today}

\begin{abstract}

 \noindent
     Non-minimally coupled  $ Y(R) $-Maxwell  gravity which  have some
 interesting solutions  may be used to understand  dark matter, dark energy, the origin of cosmic magnetic field and the evaluation of the universe.
     We give some new solutions to the model  such as spherically
symmetric electric, magnetic and dyon solutions. We  point out
 an   existence of   an electromagnetic duality transformation
for the model.



\end{abstract}

\pacs{Valid PACS appear here}
\maketitle


\def\ba{\begin{eqnarray}}
\def\ea{\end{eqnarray}}
\def\w{\wedge}



\section{Introduction}

\noindent

 \noindent  The non-minimally coupled $Y(R)$-Maxwell  models
 \cite{prasanna,horndeski,drummond,buchdahl,dereli1,muller-hoissen,balakin1,balakin2,balakin3,dereli2,lambiase,Liu,bamba1,bamba2,Sert,dereli3,dereli4,Sert2}
 have some good properties for shedding light on some concepts,
such as magnetization in matter, cosmic magnetic fields,
  dark matter,  dark energy and the evaluation of the universe.
  If one looks for a new fundamental law in order to  explain these
concepts, its best implication is to have new symmetries.
Symmetries  have  an important role in physics and mathematics.
 Because  they  imply an existence   of
a corresponding  fundamental conservation law
  and dictate the form of  the law.
 Thus the recent theoretical explorations are
mostly based on the search for new symmetries of nature. Thus,
duality symmetry of the field equations of the  $Y(R)$-Maxwell  models is the main motivation of this study.

 The duality symmetry of Einstein-Maxwell field equations has
received much attention in gauge theories and gravitation.
 It provides useful tools to obtain
new solutions to the field equations for studying different
regimes of the theory. According to this symmetry,  the
Maxwell equations $dF=0$ and $ d*F=0$ remain invariant under the
duality transformation
     $ (E,B)\rightarrow  (B, -E) $ or $ (F, *F) \rightarrow (*F, -F) $.
But,   the Maxwell Lagrangian changes sign under the transformation.
 That is, this symmetry is valid only in the  field equations, not Lagrangian.
 This duality  (Heaviside duality) of the electromagnetic
system provides a connection between  the  electric  and the
magnetic black hole solutions. Then if we know an electrically
charged solutions, we can find also corresponding
magnetically charged solutions and vice verse.
 The  duality
transformation of electromagnetic fields are  related to the
conserved quantities such as  currents and helicity.

 In this study we show that  the non-minimal theory
has  a new duality transformation which is $ (F, *F) \rightarrow
(*G, -G)  $   and $Y \rightarrow 1/Y $ (G is excitation 2-form).
In this case, similar to the minimal Einstein-Maxwell theory  the
field equations remain invariant  and the non-minimal part of the Lagrangian changes sign under the transformation,
and it becomes a different model.
While  this duality transformation provides a connection between
the electric and the magnetic black hole solutions for the
non-minimal model, additionally,  it determines the structure of
the non-minimal coupling function $Y(R)$.

In the present paper, we proceed to investigate the non-minimal
couplings of gravitational and electromagnetic fields giving a new
duality transformation for the model. We obtain some new dyon
solutions with  electric and magnetic field.
 Then, we point out that the solutions and the field equations are  left invariant under the new duality transformation.

\section{Field Equations of the Non-minimally Coupled Theory  and Duality Transformations} \label{model}

\bigskip

\noindent We obtain the field equations of the non-minimal theory by a variational
principle from an action
\begin{equation}
        I[e^a,{\omega^a}_b,F] = \int_M{L}.
        \nonumber
\end{equation}
where  $\{e^a\}$ and ${\{\omega^a}_b\}$ are the fundamental
gravitational field variables and   $F=dA$ is the electromagnetic
field 2-form.  The space-time metric $g = \eta_{ab} e^a \otimes
e^b$ has the signature $(-+++)$ and we fix the orientation by setting
$*1 = e^0 \w e^1 \w  e^2 \w e^3 $.  Torsion 2-forms $T^a$ and
curvature 2-forms $R^{a}_{\; \; b}$  are given as
\begin{equation}
T^a = de^a + \omega^{a}_{\;\;b} \w e^b , \nonumber
\end{equation}
\begin{equation}
R^{a}_{\;\;b} = d\omega^{a}_{\;\;b} + \omega^{a}_{\;\;c} \w \omega^{c}_{\;\;b} . \nonumber
\end{equation}
We start with  the following  Lagrangian density 4-form;
 \ba \label{action}
  L =  \frac{1}{2\kappa^2} R*1 -\frac{1}{2}Y(R) F\w *F +  T^a \w \lambda_a ,
   \label{Lagrange}
   \ea
where
 $\kappa^2 = 8\pi G$ is  Newton's universal gravitational constant $(c=1)$ and $R$ is the curvature scalar which can be found by applying interior product $\iota_a $ twice to the curvature tensor $R_{ab}$ 2-form.  This  Lagrangian density  involves Lagrange multiplier 2-form $\lambda_a$  whose variation imposes the zero-torsion constraint  $T^a=0$. Then we use only  the unique metric-compatible Levi-Civita
connection.
We use the shorthand notation $ e^a \wedge e^b \wedge \cdots =
e^{ab\cdots}$, and  $\iota_aF =F_a, \  \  \iota_{ba} F =F_{ab}, $ \   $ \iota_a {R^a}_b =R_b, \  \   \iota_{ba} R^{ab}= R $.
   The field equations are obtained by considering the independent variations of
   the action with respect to  $\{e^a\}$,
   ${\{\omega^a}_b\}$ and $\{F\}$.  The electromagnetic field components are read  from the expansion $F = \frac{1}{2} F_{ab} e^a  \w  e^b$.
\\
\noindent After taking the infinitesimal variations of the total Lagrangian
density $L $,
 we find the  Einstein and Maxwell field equations    \cite{dereli3,dereli4}  for the extended theory as
\begin{eqnarray}\label{einstein}
&&   \frac{1}{2 \kappa^2}  R^{bc}
\w *e_{abc} +  \frac{1}{2} Y  (\iota_a F \w *F - F \w \iota_a *F)   + Y_R  (\iota_a R^b)\iota_b( F \w *F )
 \nonumber
\\
&& + \frac{1}{2}  D [ \iota^b D(Y_R F_{mn} F^{mn} )]\wedge *e_{ab}
 =0   ,
\end{eqnarray}
\begin{equation}\label{maxwell1}
d(*Y F) = 0, \  \  \   \  dF=0.
\end{equation}
where  $Y_R = \frac{dY}{dR}$.
  One can show that the action (\ref{action})  and the field equations (\ref{einstein})-(\ref{maxwell1})
  when written out explicitly in any local coordinate system are equivalent to those given by Bamba and Odintsov\cite{bamba2}.

The non-minimal coupling model modifies both  the Maxwell and the
Einstein field equations. Modifications to the Maxwell equations
can be associated with the magnetization and the polarization of a
specific medium, while modifications to the Einstein equations
changes the space-time metric.
 Here the medium can be considered
as containing both magnetically and electrically  polarisable
matter with $G=YF$.
 Now we want to show that these field equations
(\ref{einstein})-(\ref{maxwell1}) remain invariant under the
 duality transformation $ (F, *F) \rightarrow (*G, -G)  $  and $Y
\rightarrow 1/Y $.
 As we see from the following steps, each part of the field equations
(\ref{einstein}) remain invariant;

\begin{eqnarray}
Y  (\iota_a F \wedge *F - F \wedge \iota_a *F) \rightarrow &&
(\frac{1}{Y})[\iota_a* G \w (-G) - *G \w \iota_a (-G)] \nonumber
\\ && = Y  (\iota_a F \wedge *F - F \w \iota_a *F)
\end{eqnarray}
\begin{eqnarray}
Y_R  (\iota_a R^b)\iota_b( F \wedge *F )  \rightarrow &&
(-\frac{1}{Y^2}) Y_R (\iota_a R^b)\iota_b [ *G \wedge  (-G)  ]
\nonumber
\\ && =  Y_R  (\iota_a R^b)\iota_b( F \w *F )
\end{eqnarray}

\begin{eqnarray}
\frac{1}{2}  D [ \iota^b D(Y_R F_{mn} F^{mn} )]\wedge *e_{ab}
\rightarrow && \frac{1}{2}  D [ \iota^b D((-1/Y^2) Y_R (*G)_{mn}
(*G)^{mn}
)]\wedge *e_{ab} \nonumber \\
&& = \frac{1}{2}  D [ \iota^b D(Y_R  F_{mn} F^{mn} )]\wedge
*e_{ab}
\end{eqnarray}
where $\iota_n \iota_m *G = (*G)_{mn} = Y (*F)_{mn}$. In the last
step we have used the fact that  $(*F)_{mn} (*F)^{mn} = - F_{mn}
F^{mn} $. It is obvious that the first part of (\ref{einstein})
and the Maxwell equations (\ref{maxwell1}) do not change under
this transformation. But,  the non-minimal part of    the Lagrangian density
 changes sign under the transformation
  as it is in the minimal
Einstein-Maxwell theory;

\begin{eqnarray}
 {F}\wedge {*YF} \rightarrow *G \wedge \frac{1}{Y} (-G)=-F \wedge
 *YF .
\end{eqnarray}
Thus this transformation leads to a different dual model.
\noindent If we write the explicit form of $ F $  in components as
\begin{eqnarray}
 F= E_1 e^{01} + E_2 e^{02} + E_3 e^{03} + B_1e^{23} -B_2 e^{13} +
 B_3 e^{12},
\end{eqnarray}
\noindent then we calculate $*G$
\begin{eqnarray}
 *G =  Y B_1 e^{01} +  Y B_2 e^{02} + Y B_3 e^{03} - Y E_1e^{23} + YE_2 e^{13}
 - YE_3 e^{12}.
\end{eqnarray}

\noindent Thus the transformation $ (F, *F) \rightarrow (*G, -G) $
corresponds to $({\ E_i, B_i}) \rightarrow  ({ YB_i, -YE_i}) $ in
terms of electric $E_i$ and magnetic $B_i$ components  of
electromagnetic tensor $F$,  $i=1,2,3$. Thus  we find a map
between  electric and magnetic   solutions in the electrically and
magnetically polarizable medium.


%
 \section{Some Static, Spherically Symmetric Solutions}

\noindent We consider (1+3)-dimensional static, spherically symmetric  solutions to the non-minimal model which are given by the metric
\begin{equation}\label{metric}
              g = -f(r)^2dt^2  +  f(r)^{-2}dr^2 + r^2d\theta^2 +r^2\sin(\theta)^2 d \phi^2.
\end{equation}
\subsection{Magnetic Monopole Solutions}
  \noindent In order to show   the electromagnetic duality explicitly in  solutions and in field equations,  first
  we take the electromagnetic tensor $ F$ which has only the magnetic component $B=B_1$;
 \begin{eqnarray}\label{electromagnetic1}
 F   &=& B(r) r^2  \sin(\theta) d\theta\w  d\phi=  B(r)e^2 \w e^3.
\end{eqnarray}
 The   field  equations  (\ref{einstein})  and (\ref{maxwell1}) give us   the following system of equations
    \cite{Sert}:
 \begin{eqnarray}\label{ein1}
  && \frac{1}{\kappa^2}(\frac{{f^2}\rq{}}{r}  + \frac{f^2-1 }{r^2} ) +  Y_R B^2 (\frac{ {{f^2}\rq{}}\rq{} } {2} + \frac{{f^2}\rq{} }{r}  ) + \frac{1}{2} YB^2
  + [(B^2  Y_R )\rq{} f]\rq{}f  +  \frac{2}{r} f^2 (B^2 Y_R   )\rq{}=0,
  \nonumber \\
    &&
\frac{1}{\kappa^2}(\frac{{f^2}\rq{}}{r}  + \frac{f^2-1 }{r^2} ) +  Y_R B^2 (\frac{ {{f^2}\rq{}}\rq{} } {2} + \frac{{f^2}\rq{} }{r}  ) + \frac{1}{2} YB^2
 + (B^2  Y_R )\rq{} ( \frac{{f^2}\rq{}}{2} + \frac{2 f^2}{r} ) =0 ,\\
 &&
 \frac{1}{\kappa^2}(\frac{ {{f^2}\rq{} }\rq{} }{2}  + \frac{{f^2}\rq{}}{r} ) + Y_R B^2 ( \frac{  {{f^2}\rq{}} } {r} + \frac{{f^2-1} }{r^2}  ) - \frac{1}{2} YB^2
  +  [(B^2  Y_R )\rq{} f]\rq{}f  +  (B^2  Y_R )\rq{} ( \frac{{f^2}\rq{}}{2} + \frac{f^2}{r} )=0, \nonumber\\
 && B=\frac{q_m}{r^2} \label{m1}
 \end{eqnarray}
Here the curvature scalar is calculated as
\begin{eqnarray}
R=- {{f^2}\rq{}}\rq{} -\frac{4 }{r} {f^2}\rq{} -\frac{2}{r^2} ( f^2-1) .
\end{eqnarray}

 \noindent The following  solutions to these differential equations  are
found in \cite{Sert} for some non-minimal coupling functions. That
is;
\begin{eqnarray}\label{nonminimalsol2}
f^2(r) &=& 1-\frac{2M}{r}+\frac{a_1\kappa^2q^2}{r^2}\ln \frac{r}{r_0}  +\frac{\kappa^2q^2 ( 1+ 5a_1)} {4r^2},   \hskip 1 cm for \ a_1 \neq 0,  \\
B(r) &=& \frac{q_m}{r^2},
\end{eqnarray}
   for  \begin{eqnarray}\label{nonminimalsol1}
Y(R)=1-a_1\ln( \frac{R}{R_0} )
\end{eqnarray}
and
\begin{eqnarray}\label{nonminimalsol4}
f^2(r) &=& 1-\frac{c}{r}+\frac{ \kappa^2q^2} {4r^2}  - \frac{a_2(\beta-1)^2}{4\beta(3\beta+1)}r^{\frac{2\beta+2}{\beta-1}},   \hskip 1 cm for \ R_0 \neq 0, \  \  \beta \neq 0,1,-\frac{1}{3}  \\
B(r) &=& \frac{q_m}{r^2},
\end{eqnarray}
for
\begin{eqnarray}\label{nonminimalsol3}
Y(R)=1-( \frac{R}{R_0} )^\beta
\end{eqnarray}

 \noindent where $q_m$ is  determined by the Gauss  integral
   \begin{eqnarray}
    \frac{1}{4\pi} {\int_{S^2}{ F }} =  \frac{1}{4\pi} {\int_{S^2}{ B(r) r^2 \sin \theta  d\theta \wedge d\phi}}=q_m.
    \end{eqnarray}

 \noindent To show  the existence of the electromagnetic duality in these
non-minimally coupled models,
 we compare these magnetic solutions with the previous electric solutions.  These equations we found in (\ref{ein1}) and (\ref{m1})
  for  the magnetic field    turn out to be the equations given in Ref. \cite{dereli3} (eqs. (23)-(26)) for an electric field ansatz  under these transformations
$ B\rightarrow - D_e $, $q_m \rightarrow -q_e $ and $Y\rightarrow
\frac{1}{Y}$.
  Here the displacement field is defined  $D_e=YE $ (see Ref. \cite{dereli3,dereli4} taking $G=YF=YEe^{01} $ ).
Consequently the above  solutions   turn out to be the solutions
given in Ref. \cite{dereli3,dereli4}. Thus we reach the previous different model.


\subsection{Dyon Solutions }

\noindent In this subsection our aim is to find new solutions and
to show their electromagnetic duality for the non-minimal
model. Now we consider  the electromagnetic field of a  dyon which
has the magnetic and electric components;
 \begin{eqnarray}\label{electromagneticdyon}
 F   &=& E(r) d t \w d r +  B(r) r^2  \sin(\theta) d\theta\w  d\phi= E(r)e^0 \w e^1 +  B(r)e^2 \w e^3.
\end{eqnarray}
We calculate the   field  equations for the metric (\ref{metric})
together with the dyon electromagnetic field 2-form
(\ref{electromagneticdyon}). Then we find
\begin{eqnarray}\label{dif3}
\frac{1}{\kappa^2}\left( {{f^2}\rq{}}\rq{} -\frac{2}{r^2}(f^2-1) \right)  -  YE^2 - YB^2 &=&0,\\
YE &=& \frac{q_e}{r^2},\\
B &=& \frac{q_m}{r^2}.
\end{eqnarray}
under the constrain
\begin{eqnarray}\label{dif4}
Y_R (E^2-B^2)=\frac{1}{\kappa^2},
\end{eqnarray}
 We emphases that  these differential equations (\ref{dif3})-(\ref{dif4}) are invariant under the duality transformations
 $({ E, B}) \rightarrow  ({ YB, -YE}) $, $(q_e,q_m ) \rightarrow ( q_m,-q_e)$ and $Y \rightarrow 1/Y $.

Additionally, one can look for a non-minimal function for a useful
 gravitational metric.
Now we consider the following  metric function with an extra logarithmic term
\begin{eqnarray}
f^2(r)=1- \frac{C_1}{r}  - C_3 ln(r)
\end{eqnarray}
which is  viable  to obtain the constant velocity of the trajectories  for large values of the radial distance  \cite{Fabris}. They also comes  from the computation of a running gravitational coupling as a result of the quantum effects \cite{Shapiro} and the TeVeS theory
\cite{Bekenstein,Bruneton}. Here $C_1$ and $C_3$ are constants. Then we determine the structure of the non-minimal function $Y$
which is consistent with the solution.
 After solving the  field equations for the model,
we find  the non-minimal function, the electric and magnetic
fields;
\begin{eqnarray}
Y(r) &=& \frac{C_3}{2q_m^2 \kappa^2}r^2 + \frac{C_3}{q_m^2\kappa^2}r^2ln(r) \nonumber  \\
&&  -\frac{1}{2 q_m^2\kappa^2 } \sqrt{     C_3^2r^4  +  4C_3^2 r^4 ln(r)(1+ln(r))  -   4\kappa^4q_e^2 q_m^2  }\\
B(r)&=& \frac{q_m}{r^2}\\
E(r) &= & \frac{q_e}{Y(r) r^2}.
\end{eqnarray}
We note that the dependence of $Y$ to $R$ is  not explicit.

\noindent In order to find
  asymptotically flat solutions with an  explicit $Y(R)$ function,  we consider
\begin{eqnarray}
f^2(r)=1- \frac{2M}{r} + \frac{C_2}{r^2} -\frac{C_3 ln(r)}{r },
\end{eqnarray}
then  for $C_3\neq 0$ we obtain
\begin{eqnarray}
Y(r) &=& \frac{2C_2}{q_m^2 \kappa^2} + \frac{3C_3}{2q_m^2\kappa^2}r -\frac{1}{2 q_m^2\kappa^2 } \sqrt{      16C_2^2  -  4q_e^2\kappa^4q_m^2  +   24C_2C_3 r + 9C_3^3r^2    }\\
B(r)&=& \frac{q_m}{r^2}\\
E(r) &= & \frac{q_e}{Y(r) r^2}.
\end{eqnarray}
We can solve $r$ form $R(r)=\frac{C_3}{r^3}$, thus we can find $Y(R)$ :
\begin{eqnarray}
Y(R) =\frac{2C_2}{q_m^2 \kappa^2} + \frac{3C_3^{\frac{4}{3}}}{2q_m^2\kappa^2}R^{-\frac{1}{3} } -\frac{1}{2q_m^2\kappa^2}\sqrt{         16C_2^2 -4q_e^2\kappa^2 +    24C_2C_3^{\frac{4}{3}} R^{-\frac{1}{3}} + 9C_3^{\frac{8}{3}}  R^{-\frac{2}{3}}   }.
\end{eqnarray}

\noindent Lastly, we can find   the solutions with an extra correction  power  law term;
\begin{eqnarray}
f^2(r)&=&1 + \frac{C_1}{r}  - C_3 r^n.
\end{eqnarray}
which  is  a candidate to solve both the dark energy and dark matter problem \cite{Capozziello}.
\begin{eqnarray}
Y(r) &=& \frac{C_3}{2q_m^2 \kappa^2}(n^2-n- 2)r^{n+2}\nonumber\\
&& - \frac{1}{2q_m^2\kappa^2}  \sqrt{     C_3^2 ( n^4 - 2 n^3 - 3n^2 +4n +4 ) r^{2n+4}    -   4\kappa^4q_e^2 q_m^2  }\\
B(r)&=& \frac{q_m}{r^2}\\
E(r) &= & \frac{q_e}{Y(r) r^2}
\end{eqnarray}
Here we note that  one can consider the case  $q_e=0$ in these
solutions. Thus one can reach  magnetically charged solutions.
This means that we have also corresponding electrically charged
solutions because of the duality transformation.

\section{Conclusion}

\noindent
 We  have investigated   the duality transformation   of the
$Y(R)$-Maxwell theory of gravity and  new solutions of it.
  We find firstly a new duality transformation  which   leave invariant
   the field equations, while  the non-minimal part of the Lagrangian
changes sign under that transformation, and this leads to a  different model.
 This duality transformation is a map
 between the electromagnetic field $ F $ with
$*G=*Y(R)F $
 under which all the non-minimal gravitational solutions
are mapped into new solutions. Also it determines a relation about
how the non-minimal coupling  function $Y(R)$ transforms in the
specific medium. Thus we can find new solutions and categorize
them in order to gain more insights on the electric and magnetic
fields in the medium.
 We give  some  examples to the duality which transforms
spherically symmetric, magnetic solutions to the electric
solutions in the electrically and magnetically polarizable medium.
 After then  we find various non-minimal
functions of the theory for some interesting spherically symmetric
metric functions
\cite{Fabris,Shapiro,Bekenstein,Bruneton,Capozziello}
 with the magnetic  and electric charge.
Finally we  show that these solutions also satisfy this duality.

\vskip 1cm

\section{Acknowledgement}

\noindent
This work was supported by  Pamukkale University (BAP  Project No: 2012BSP014).


\vskip 1cm



\section{References}

\begin{thebibliography}{99}



 \bibitem{prasanna} A. R. Prasanna, {\it Phys. Lett. }  A {\bf  37}, 337 (1971)


\bibitem{horndeski}  G. W. Horndeski, {\it J. Math. Phys.} {\bf 17}, 1980
(1976)

\bibitem{drummond}  I. T. Drummond, S. J. Hathrell, {\it Phys. Rev. D} {\bf 22}, 343
(1980)

\bibitem{buchdahl} H. A. Buchdahl, {\it J. Phys. A } {\bf 12}, 1037
(1979)
\bibitem{dereli1} T. Dereli, G. \"{U}\c{c}oluk, {\it Class. Q. Grav.} {\bf 7}, 1109 (1990).
\bibitem{muller-hoissen} F.  M\"{u}ller-Hoissen, {\it Class. Q. Grav.} {\bf 5}, L35
(1988)


\bibitem{balakin1}  A. B. Balakin,  J. P. S. Lemos,  {\it Class. Q. Grav.}  {\bf 22}, 1867
(2005)
\bibitem{balakin2}  A.B. Balakin, H. Dehnen,  A.E. Zayats,  {\it Phys. Rev. D} {\bf  79}
  024007, (2009)

 \bibitem{balakin3}  A.B. Balakin and A.E. Zayats, {\it Gravitation and Cosmology }
 {\bf 14},  86-94,  (2008)


 \bibitem{dereli2} T. Dereli, \"{O}. Sert, { \it Phys. Rev. D  } {\bf 83}, 065005 (2011)

 \bibitem{lambiase} G. Lambiase, S. Mohanty and G. Scarpetta,   {\it  JCAP} {\bf  07}, 019, (2008)

 \bibitem{Liu} Y. Liu and J. Jing,  {\it Gen. Relativ. Gravit.}, {\bf  44}, 7 (2012)




\bibitem{bamba1}  K. Bamba and S. D. Odintsov,    {\it JCAP} {\bf 0804},  024, (2008)
\bibitem{bamba2}  K. Bamba,  S. Nojiri and S. D. Odintsov,    {\it JCAP}, {\bf 0810}, 045, (2008)

 \bibitem{Sert}   \"{O}. Sert,   {\it Eur. Phys. J. Plus } 127: 152 (2012)

  \bibitem{dereli3} T. Dereli, \"{O}. Sert, {\it Eur. Phys. J. C}  {\bf 71}, 3, 1589 (2011)

   \bibitem{dereli4} T. Dereli, \"{O}. Sert,   { \it Mod. Phys. Lett. A}  {\bf  26}, 20, 1487-1494 (2011)
\bibitem{Sert2} \"{O}. Sert and M. Adak, "An anisotropic cosmological solution to the Maxwell-$Y(R)$
gravity",     arXiv:1203.1531 [gr-qc]


  \bibitem{Fabris}  J.C. Fabris, J. P. Campos   {\it Gen. Relativ. Gravit.}, {\bf 41}, 1 (2009)

\bibitem{Shapiro}  I.L. Shapiro, J. Sola and H. Stefancic,  {\it JCAP}, {\bf 01}, 012 (2005)

\bibitem{Bekenstein} J. Bekenstein, {\it Phys. Rev. D } {\bf 70},  3509 (2004)

\bibitem{Bruneton}  J.Ph.,  Bruneton, G.,  Esposito-Farese,  {\it Phys. Rev. D} {\bf 76}, 124012 (2007)

\bibitem{Capozziello}  S., Capozziello, V.F., Cardone, A. Troisi, {\it Mon. Not. Roy. Astron. Soc. }, {\it  375} 4 (2007)



\end{thebibliography}
\end{document}